\documentclass[12pt]{article}
\usepackage{latexsym}
\usepackage{amssymb, amsmath, xspace, lscape,  latexsym}

\newcommand{\bea}{\begin{eqnarray}}
\newcommand{\eea}{\end{eqnarray}}
\def\beq#1#2\eeq{
        \begin{equation}
        \label{#1}
            #2
        \end{equation}}

\newcommand{\al}{\alpha}
\newcommand{\bt}{\beta}

\renewcommand{\tilde}{\widetilde}

\newcommand{\bq}{\begin{eqnarray}}
\newcommand{\eq}{\end{eqnarray}}
\newcommand{\nn}{\nonumber}
\newcommand{\ba}{\begin{array}}
\newcommand{\ea}{\end{array}}

\def\btheor#1\etheor{
        \begin{theor}
            #1
        \end{theor}
    }

    \def\bsled#1\esled{
        \begin{sled}
            #1
        \end{sled}   }

\newcommand{\la}{\lambda}

\def\hm#1{#1\nobreak\discretionary{}{\hbox{\m@th$#1$}}{}}
\def\mi#1{\discretionary{\hbox{\m@th$#1$}}{\hbox{\m@th$#1$}}{}}

\vspace{4ex}

\newcommand{\rme}{{\rm e}}

\newcommand{\qed}{\hfill $\Box$}

\newcommand{\be}{\begin{equation}}
\newcommand{\ee}{\end{equation}}
\newcommand{\bW}{{\bf W}}

\newcommand{\bI}{{\bf I}}
\newcommand{\bH}{{\bf H}}
\newcommand{\bx}{{\bf x}}
\newcommand{\by}{{\bf y}}

\newcommand{\bxd}{{\bf x}^{\dag}}
\newcommand{\bHd}{{\bf H}^{\dag}}
\newcommand{\bn}{{\bf n}}

\renewcommand{\v}{{\mathsf{v}}}

\newcommand{\vp}{{\mathsf{v}^{\prime}}}

\newcommand{\dN}{{\delta_N}}
\newcommand{\tD}{\tilde{D}}

\numberwithin{equation}{section}

\begin{document}

\title{\bf Perturbed Laguerre Unitary Ensembles, Hankel Determinants, and Information Theory}
\author{Estelle Basor \\
        {\small American Institute of Mathematics
        (ebasor@aimath.org)} \\
        Yang Chen \\
        {\small Department of Mathematics, University of Macau (yayangchen@umac.mo)} \\
        Matthew R. McKay \\
        {\small ECE Department, Hong Kong University of Science and
        Technology (eemckay@ust.edu)}
        }

\date{}

\maketitle

\begin{abstract}
This article investigates a
key information-theoretic performance metric in multiple-antenna
wireless communications, the so-called outage probability.  The article is partly a review, with the methodology based mainly on \cite{ChenMckay}, whilst also presenting some new results. The
outage probability may be expressed in terms of a moment generating function, which involves a Hankel determinant generated from a perturbed Laguerre weight.
For this Hankel determinant,  we present two separate integral representations, both involving solutions to certain non-linear differential equations. In the second case, this is identified with a particular $\sigma$-form of Painlev\'e V. As an alternative to the Painlev\'e V, we show that this second integral representation may also be expressed in terms of a non-linear second order
difference equation.
\end{abstract}

\section{Introduction}

In Random Matrix Theory (RMT) many statistical quantities can be
described as determinants. This is especially true in the cases of
what are known as the classical ensembles, for example, the Gaussian
Unitary Ensemble (GUE) or the Laguerre Ensembles. Since the
groundbreaking work of Tracy and Widom \cite{TW1994}, which
characterized the largest eigenvalue distribution for the GUE, one
common approach is to write some statistical quantity as a
determinant and then express the determinant as something involving
a Painlev\'e transcendent, a solution to one of the classical
Painlev\'e second order non-linear differential equations.

Techniques to find the determinant and then the resulting
differential equation are quite complicated. Tracy and Widom used
more of an operator theory approach, others have used an integrable
system approach, whilst others still have used a stochastic equation
approach.

This paper highlights a technique known as the ``ladder operator''
approach and is, effectively, an example illustrating the
technique and describing the application of interest.  The specific application relates to the performance of multiple-input
multiple-output (MIMO) wireless communication systems, in which both
the transmitter and receiver devices are equipped with multiple
antennas. Such systems have been the subject of intense interest
since the key papers \cite{Telatar} and \cite{Foschini}, and now
form the cornerstone of most modern day wireless systems (Wi-Fi,
cellular networks, etc). Here, a key fundamental performance measure
is considered, the so-called ``outage probability'',  and this is
shown to involve the probability distribution of a certain ``linear
statistic'' in a Laguerre random matrix ensemble. The problem thus
falls naturally within the realm of the ladder operator framework.
We should point out that while we emphasize the usefulness of the 
technique with the application to communication systems, this technique
has been successful in many other settings to relate some statistical quantity to a Painlev\'e tnascendent. In particular, the 
papers \cite{Basor, BCN, BCZ, chen+its, CN, ch-zh, ChenMckay} all use the ladder operator approach illustrated here.

Here is the idea of the approach. For the example at hand, the
linear statistic of interest can be described through its moment
generating function as a Hankel determinant, which using the theory
of orthogonal polynomials (for a certain nonstandard weight
generally) can be computed via a product of norms of monic
orthogonal polynomials. Now it is well-known that orthogonal
polynomials satisfy three term recurrence equations. In fact, the
two coefficients in the recurrence combined with initial conditions,
completely determine the polynomials. Thus information about the
coefficients in the recurrence equations should yield information
about the Hankel determinants.

The path to this information is from the ladder operators, two
formulas that connect the polynomials one index apart to their
derivatives. These yield, using basic complex analysis, a set of
equations in the coefficients along with two additional auxiliary
quantities that arise.  The story would end here, except for the
fact that often there is a ``time'' parameter implicit in the
original weight and thus in the polynomials themselves. Using
``time'' evolution one can then, using only elementary means, find a
pair of coupled Ricatti equations in the two auxiliary quantities.
These then lead directly to a Painlev\'e equation. It should be
pointed out that this method works at least in principle, if the
``time'' parameter is present and if the derivative of the logarithm
of the weight is a rational function. Then one can in many cases
follow the steps illustrated in this paper.

Here is an outline of the paper. The next section contains the
preliminaries of the theory including the ladder operator equations.
Section III shows how the application of interest, the outage
probability performance measure which arises in the application of
MIMO wireless communication, can be described using the RMT
framework. Section IV provides the details of the path to the
differential equation solutions, which are presented Theorems 1 and
2.

We point out that the aim of this paper is to give an expository review of the ladder operator approach, largely following the developments in \cite{ChenMckay}; however, some new results are also presented.  In particular, these pertain to the result in Theorem 1, and also the discrete $\sigma$-form relation in Theorem 2.

\section{Preliminaries}

\subsection{Linear Statistics of Hermitian Random Matrices and Hankel Determinants}

For the MIMO capacity application, it will be seen that the problem
of interest falls within the general theory of linear statistics of
Hermitian random matrices, with a close connection to the theory of
orthogonal polynomials. Here a brief introduction to the general
theory is given, and preliminaries are established for later use.

We will require the distribution of a certain linear statistic
\begin{align} \label{eq:LinStat}
\sum_{k=1}^N f(x_k)
\end{align}
in the eigenvalues $\{ x_k \}$ of a $N \times N$ Hermitian random
matrix, with joint eigenvalue density of the form
\begin{align} \label{eq:EigDensGeneric}
p(x_1, \ldots, x_N) \propto \prod_{k = 1}^N  w_0(x_k) \prod_{1 \leq
i <  j \leq N} ( x_j - x_i )^2  ,
 \quad  x_k \in (a, b)
\end{align}
for some weight function $w_0(\cdot)$. It is convenient to attempt
to characterize the distribution of the linear statistic
(\ref{eq:LinStat}) through its moment generating
function\footnote{The parameter $\lambda$ is an indeterminate which
generates the random variable
$\sum_{k=1}^N f(x_k).$\\
},
\begin{align} \label{eq:MGFFirst}
\mathcal{M}(\la) = E \left[ \exp \left( \la \sum_{k=1}^N f (x_k)
\right) \right] \; = \; E \left[ \prod_{k=1}^N e^{\la f(x_k)}
\right]
\end{align}
which upon substituting for (\ref{eq:EigDensGeneric}) gives
\begin{align} \label{eq:DnGeneric}
\mathcal{M}(\la) &= \frac{ \frac{1}{N!} \int_{(a,b)^N}\prod_{1\leq
i<j\leq N}(x_j-x_i)^2\prod_{k=1}^{N} w(x_k) dx_k} {\frac{1}{N!}
\int_{(a,b)^N}\prod_{1\leq i<j\leq
N}(x_j-x_i)^2\prod_{k=1}^{N}w_0(x_k) dx_k}
\end{align}
where
$$
w(x) := w_0(x) e^{\la f(x) }
$$
denotes the \emph{deformed} version of the reference weight
$w_0(\cdot)$. Application of the Andreief-Heine identity \cite{Sze}
now directly leads to
\begin{align} \label{eq:MGFRatio}
\mathcal{M}(\la) = \frac{D_N[ w ]}{D_N[w_0 ]} \; = \; \frac{ \det
\left( \int_a^b x^{i+j-2} w(x) dx \right)_{i,j=1}^{N} } {\det \left(
\int_a^b x^{i+j-2}w_0(x) dx \right)_{i,j=1}^{N}}\; \; ,
\end{align}
which is a ratio of Hankel determinants. By virtue of the Selberg
integral, for most ``classical'' weight functions of interest, the
Hankel determinant in the denominator of (\ref{eq:DnGeneric}) admits
an explicit closed-form (non determinantal) representation. The
numerator, on the other hand, is much more difficult to
characterize, since it involves the more complicated deformed weight
$w$. To proceed, methods based on orthogonal polynomials will be
introduced in the sequel.

\subsection{Orthogonal Polynomials and their Ladder Operators}
\label{sec:LadderOperators}

We start by noting that
\begin{align}
\prod_{1\leq i < j \leq N}(x_j-x_i) = \det \left( x_j^{i-1}
\right)_{i,j=1}^N = \det \left( P_{i-1}(x_j) \right)_{i,j=1}^N
\end{align}
where $P_j (\cdot)$ represents any monic polynomial of degree $j$,
 {\small \bq \label{eq:MonicDefn}
P_j(z)=z^{j}+\textsf{p}_1(j)\:z^{j-1}+... \eq} Applying this in
(\ref{eq:DnGeneric}) and once again integrating via the
Andreief-Heine identity, the numerator evaluates to
\begin{align} \label{eq:DnIntermed}
D_N[w] = \det \left( \int_a^b P_{i-1}(x) P_{j-1}(x) w(x) dx
\right)_{i,j=1}^{N} \; .
\end{align}
If we orthogonalize the polynomial sequence $\{P_{n}(x)\}$ with
respect to $w(x)$ over the interval $[a,b],$ i.e.,
\begin{align}
\label{eq:OrthRel} \int_{a}^{b}\:P_{j}(x)P_k(x)
w(x)dx=h_{j}\delta_{j,k},\;\; j,k=0,1,2,...
\end{align}
with $h_j$ denoting the square of the $L^2$ norm of $P_j$ over
$[a,b],$ then (\ref{eq:DnIntermed}) reduces to
\begin{align} \label{eq:DetRelation}
D_N[ w ] = \prod_{k=0}^{N-1} h_k \; .
\end{align}

The key challenge is how to characterize the class of polynomials
which obey the orthogonality constraints in (\ref{eq:OrthRel}) or,
more importantly, the norms of such polynomials required to evaluate
(\ref{eq:DetRelation}).


If all the moments of the weight $w$ exist, then the theory of
orthogonal polynomials states that the $P_n(z)$ for $n=0,1,2,...$
satisfy the three term recurrence relations,
\begin{align}
\label{eq:RecRel}
zP_n(z)=P_{n+1}(z)+\al_n\:P_n(z)+\bt_n\:P_{n-1}(z).
\end{align}
The above sequence of polynomials can be generated from the
orthogonality conditions (\ref{eq:OrthRel}), the recurrence
relations (\ref{eq:RecRel}), and the initial conditions,
\begin{align}
P_0(z)=1,\;\;\bt_0P_{-1}(z)=0.
\end{align}
Substituting (\ref{eq:MonicDefn}) into the recurrence relations, an
easy computation shows that \bq \label{eq:p1Defn}
\textsf{p}_1(n)-\textsf{p}_1(n+1)=\al_n, \eq with
$\textsf{p}_1(0):=0.$ A telescopic sum of (\ref{eq:p1Defn}) gives
\begin{align}
\textsf{p}_1(n) =-\sum_{j=0}^{n-1} \al_j .
\end{align}
>From the recurrence relation (\ref{eq:RecRel}) and the orthogonality
relations (\ref{eq:OrthRel}), we find {\small \bq\label{eq:betah}
 \bt_n=\frac{h_n}{h_{n-1}}. \eq}

We shall see that $\textsf{p}_1(n)$ plays an important role in later
developments. For more information on orthogonal polynomials, we
refer the reader to Szeg\"o's treatise \cite{Sze}.

Next, we present three Lemmas which are concerned with the ``ladder
operators'' associated with orthogonal polynomials, as well as
certain supplementary conditions. Note that these have been known
for quite sometime; we reproduce them here for the convenience of
the reader using the notation of \cite{chen+its}, where one can also
find a list of references to the literature. We also mention that
Magnus \cite{Magnus} was perhaps the first to apply these
lemmas---albeit in a slightly different form---to random matrix
theory and the derivation of Painlev\'e equations. Tracy and Widom
also made use of the compatibility conditions in their systematic
study of finite $n$ matrix models \cite{twdet}. See also
\cite{for1}.
\\
\noindent {\bf Lemma 1} \emph{Suppose $\v(x)=-\log w(x)$ has a
derivative in some Lipshitz class with positive exponent. The
lowering and raising operators satisfy the differential-difference
formulas:} {\small \bq
P_n'(z)&=&-B_n(z)P_n(z)+\bt_n\:A_n(z)P_{n-1}(z)\\
P_{n-1}'(z)&=&[B_n(z)+\vp(z)]P_{n-1}(z)-A_{n-1}(z)P_n(z), \eq} where
{\small \bq
A_n(z)&:=&\frac{1}{h_n}\int_{a}^{b}\frac{\vp(z)-\vp(y)}{z-y}\:P_n^2(y)w(y)dy \label{eq:AnNewDefn}\\
B_n(z)&:=&\frac{1}{h_{n-1}}\int_{a}^{b}\frac{\vp(z)-\vp(y)}{z-y}P_n(y)P_{n-1}(y)w(y)dy.
\label{eq:BnNewDefn} \eq} A direct computation produces two
fundamental supplementary or compatibility conditions valid for all
$z\in \mathbb{C}\cup\{\infty\}$. These are stated in the next Lemma.
\\
{\bf Lemma 2} \emph{The functions $A_n(z)$ and $B_n(z)$ satisfy the
supplementary conditions:} {\small $$
B_{n+1}(z)+B_n(z)=(z-\al_n)A_n(z)-\vp(z)\eqno(S_1)
$$
$$
1+(z-\al_n)[B_{n+1}(z)-B_n(z)]=\bt_{n+1}A_{n+1}-\bt_nA_{n-1}(z).\eqno(S_2)$$}
\\
It turns out that there is an equation which gives better insight
into the coefficients $\al_n$ and $\bt_n$, if $(S_1)$ and $(S_2)$
are suitably combined to produce a ``sum rule'' on $A_n(z).$ We
state this in the next lemma. The sum rule, we shall see later,
provides important information about the logarithmic derivative of
the Hankel determinant.
\\
{\bf Lemma 3} \emph{The functions $A_n(z),$ $B_n(z)$, and the sum}
{\small $$ \sum_{j=0}^{n-1}A_j(z),
$$}
\emph{satisfy the condition:} {\small $$
B_n^2(z)+\vp(z)B_n(z)+\sum_{j=0}^{n-1}A_j(z)=\bt_n\:A_n(z)\:A_{n-1}(z).\eqno(S_2')
$$}

\section{Information Theory of MIMO Wireless Systems}

In this section we introduce the wireless communication problem of
interest, and connect it with the general linear statistics
framework introduced previously.

We consider a MIMO communication system in which a transmitter
equipped with $n_t$ antennas communicates with a receiver equipped
with $n_r$ antennas. Denoting the transmitted signal vector as $\bx
\in \mathbb{C}^{n_t}$ and the received signal vector as $\by \in
\mathbb{C}^{n_r}$, under a certain assumption on the channel (known
as ``flat fading''), these signals are related via the linear model
{\small \bq \label{eq:LinearModel} \by = \bH \bx + \bn \; , \eq}
where $\bn  \in \mathbb{C}^{n_r}$, the receiver noise vector, is
complex Gaussian with zero mean and covariance $E( \mathbf{n}
\mathbf{n}^\dagger ) = \mathbf{I}_{n_r}$. The \emph{channel
matrix}, $\bH \in \mathbb{C}^{n_r \times n_t}$,
 represents the wireless fading coefficients
between each transmit and receive antenna. The channel is modeled
stochastically, with distribution depending on the specific wireless
environment. Under the assumption that there are
sufficient scatterers surrounding the transmit and receive
terminals, the channel matrix $\bH$ is well modeled by a
complex Gaussian distribution with independent and identically
distributed (i.i.d.) elements having zero mean and unit variance.
This matrix is assumed to be known at the receiver\footnote{In
practice, this information can be obtained using standard estimation
techniques.}, but the transmitter only has access to its
distribution. The transmitted signal $\bx$ is designed to meet a
power constraint, $E(\bxd\bx)\leq P$. 

Our objective is to study the fundamental capacity limits of a MIMO
communication system. Such limits are described by the field of
information theory, founded by Claude Shannon in 1948
\cite{Shannon}. Specifically, information-theoretic measures allow
one to precisely determine the highest data rate that can be
communicated with negligible errors by any transmission scheme.
Consequently, information theory offers a benchmark for the design
of practical transmission technologies, and has become an
indispensable tool for modern communication system design.

The capacity of a communication link is determined by the so-called
``mutual information'' between the input and output signals. For the
MIMO model (\ref{eq:LinearModel}) it is given by: {\small \bq
\label{eq:MIDefn}
I(\bx;\by | \bH )
={\cal H}(\by | \bH)-{\cal H}(\bn) \eq } with ${\cal H}(\by | \bH)$
denoting the conditional entropy of $\by$, {\small \bq {\cal H}(\by
| \bH)= -\int_{\mathbb{C}^{n_r} }p(\by | \bH)\log p(\by | \bH)d \by
, \eq} where $p(\by | \bH)$ denotes the conditional density of $\by$
given $\bH$.  This formula represents the maximum amount of
information that can be reliably transported between the transmitter
and receiver (i.e., it represents the rate which is ``supportable''
by a given realization of the MIMO channel). It was proved in
\cite{Telatar} that the conditional mutual information $I(\bx;\by |
\bH )$ is maximized by choosing the input signal vector $\bx$
according to a zero-mean circularly-symmetric complex Gaussian
distribution with covariance ${\bf Q}_{\rm x} = E \left( \bx \bxd
\right)$ satisfying ${\rm tr} \left({\bf Q}_{\rm x} \right) \leq P$.
In this case, the mutual information (\ref{eq:MIDefn}) was shown to
be {\small \begin{align}
 \label{eq:MIDefn2} I(\bx;\by | \bH ) = \log \det \left( \bI_{n_r}  + \bH {\bf Q}_{\rm x} \bHd
 \right) \; .
\end{align}}

In this paper, we will consider a scenario in which the channel is
selected randomly at the beginning of a transmission, and remains
\emph{fixed} during the transmission.  In this scenario, it is
impossible to guarantee that the communication will be completely
reliable, since no matter what transmission rate $R$ we choose
(which is assumed fixed) there is always a non-zero probability that
the rate may not be supportable by the channel. In other words,
there is always a chance that the mutual information $I(\bx;\by |
\bH )$ falls below $R$, and thus communicating at rate $R$ becomes
impossible. This is referred to as an ``outage event'', and the
probability of this occurring is called the \emph{outage
probability}, {\small \begin{align} \label{eq:outProb} P_{\rm out} (
R ) &:= {\rm Pr} \left( I(\bx;\by | \bH ) < R  \right)
.
\end{align}}


Here, we will make the common assumption that {\small \begin{align}
\label{eq:xCov} \mathbf{Q}_{\rm x} = \frac{P}{n_t} \bI_{n_t} \; ,
\end{align}}
corresponding to sending independent complex Gaussian signals from
each transmit antenna, each with power $P/n_t$. Hence, with this
input signal covariance, the quantity $P \, (>0)$ will also
represent the signal-to-noise ratio (SNR).  With $\mathbf{Q}_{\rm
x}$ given by (\ref{eq:xCov}), the mutual information $I(\bx;\by |
\bH )$ becomes
{\small \begin{align} I(\bx;\by | \bH ) = \log\det \left( \bI_{n_r}
+ \frac{1}{t} \bH \bHd \right) , \; \quad t := \frac{n_t}{P} .
\end{align}}
To fix notation, let $M := \textsf{max}\{n_{r},n_{t}\}, N :=
\textsf{min}\{n_{r},n_{t}\} ,\alpha := M - N$ and define {\small \bq
\bW &:=& \left\{
\begin{array}{lr}
 \bH\bHd,  &  n_{r}<n_{t}\nn\\
\bHd\bH, &    n_{r} \geq n_{t} \nn
\end{array}
\right. \eq} with positive eigenvalues $\{x_{k}\}_{k=1}^{N}$. With
these definitions, and with $\det(\bI+ {\bf A} {\bf
B})=\det(\bI+{\bf B}{\bf A})$, we can further evaluate {\small
\begin{align} \label{eq:linStat}
I(\bx;\by | \bH ) &= \sum_{k=1}^N \log \left( 1 + \frac{1}{t} x_k
\right)  \;  \nonumber \\
&= - N \log t + \sum_{k=1}^N \log \left( t + x_k \right) .
\end{align}}

Computation of the outage probability (\ref{eq:outProb}) requires
the probability distribution of $I(\bx;\by | \bH )$. From
(\ref{eq:linStat}), this is clearly a linear statistic in the
eigenvalues of the Hermitian random matrix $\bW$ (with a constant
shift of $-N \log t$). Moreover, $\bW$ is complex Wishart
distributed \cite{Muirhead}, thus the eigenvalues
$\{x_{k}\}_{k=1}^{N}$ are well-known to admit the joint density
{\small
\begin{align}
 \label{eq:EigPDF_Wishart} p(x_1,x_2,...,x_N)\propto
\prod_{l=1}^{N} w_{{\rm Lag}} (x_l) \prod_{1\leq j<k\leq
N}(x_k-x_j)^2 , \quad \quad  x_l\in[0,\infty)
\end{align}
} where {\small
\begin{align} w_{{\rm Lag}} (x):=x^{\al}\:\rme^{-x},
\;\;\al>-1,\;\;x\in[0,\infty)
\end{align}}
is the classical Laguerre weight.

Our aim will be to compute the moment generating function of the
linear statistic, {\small
\begin{align}
{\tilde {\cal M}}(\lambda) := E_{\bH} \left( e^{\lambda I(\bx;\by |
\bH) } \right)  \; = \; t^{-N \la} {\cal M}(\lambda)
\end{align}
} where ${\cal M}(\lambda)$ is identified by (\ref{eq:MGFFirst}) but
with the following particularizations:
\begin{align}
\left( f(x), w_0(x), w(x), a, b \right) \Longrightarrow \left(
\log(t+x), w_{\rm Lag}(x), w_{\rm dLag}(x, t), 0, \infty \right) \;
\end{align}
where $w_{\rm dLag}(x,t)$ is a deformed Laguerre weight, {\small
\begin{align}
\label{eq:DefLagWeight} w_{\rm dLag}(x,t):= (x+t)^{\la} w_{{\rm
Lag}} (x) ,\;\;\; t>0 \;\;\;x>0 \; .
\end{align}
}
Thus, (\ref{eq:MGFRatio}) immediately gives {\small
\begin{align}
 \label{eq:MGF_MainResultsSec} {\cal M}(\lambda) = \frac{D_N(t,\la)}{D_N[w_{\rm Lag}]} \;
\end{align}}
where \bq D_N(t,\la)=\det\left( \mu_{i+j-2}(t, \la)
\right)_{i,j=1}^{N} \label{eq:HankelDefin1} \eq is the Hankel
determinant generated from $w_{\rm dLag}(x,t)$ with moments {\small
\begin{align} \label{eq:MomentDefn}
 \mu_{k}(t, \la) := \int_{0}^{\infty}x^{k} w_{\rm dLag}(x) dx \;
, \hspace*{1cm} k = 0, 1, 2, \ldots .
\end{align}
}


The quantity $D_N[w_{{\rm Lag}}]$ in the denominator of
(\ref{eq:MGF_MainResultsSec}) is the Hankel determinant generated
from the classical Laguerre weight, $w_{{\rm Lag}} (x)$, and can be
computed in terms of the Barnes $G$--function as {\small
\begin{align} \label{eq:DnLaguerre_LambdaZero} D_N[w_{{\rm
Lag}}]=\frac{G(N+1)G(N+\al+1)}{G(\al+1)},\quad G(1)=1.
\end{align}}

Our next objective will be to compute a non-determinantal
representation for the (scaled) moment generating function
(\ref{eq:MGF_MainResultsSec}).  This, in turn, will require
evaluation of the Hankel determinant $D_N(t,\la)$ in
(\ref{eq:HankelDefin1}).  We will address this problem in the sequel
by appealing to the orthogonal polynomial framework introduced in
Section \ref{sec:LadderOperators}.  We should like to mention here
that, unlike the classical ladder operators, the ``coefficients'' in
our ladder operators are $``x"$ dependent, as we shall see later.  We will present effectively \emph{three} equivalent representations, which are summarized in theorems below.

\setcounter{equation}{0}




\section{Integral Representations for the Hankel Determinant}

\subsection{Main Results}

{\bf Theorem 1\;\;} \emph{The Hankel determinant $D_N(t,\lambda)$
admits the following integral representation:}
\begin{align}
\label{eq:Intrep} \frac{D_N(t,\lambda)}{D_N[w_{\rm Lag}]} =
t^{N\lambda} \exp \left( \int^{\infty}_{t}f(y(s),y'(s),s) \frac{ds}{s} \right)
\end{align}
\emph{where}
\begin{align} &f(y(s),y'(s),s):=
\frac{\lambda^2+2(s+\alpha-\lambda)y+(4 N
s+(s+\alpha)^2-2(s+2\alpha)\lambda+\lambda^2)y^2
}{4y(y-1)^2} \\
& \hspace*{4cm} + \frac{-2(2 N
s+\alpha(s+\alpha-\lambda))y^3+\alpha^2y^4-[s\;y'(s)]^2}{4y(y-1)^2}
\; .
\end{align}

\noindent {\bf Theorem 2\;\;} \emph{The Hankel
determinant $D_N(t,\lambda)$ also admits the following equivalent integral
representation:} \bq \label{eq:HankelDet_ContSigma}
\frac{D_N(t,\lambda)}{D_N[w_{\rm Lag}]}= t^{N\la} \exp
\left(\int_{\infty}^t \frac{H_N(x)-N\la}{x} dx \right)
\label{eq:tmp} \eq \emph{where} $H_N(t)$ \emph{ satisfies the
Jimbo-Miwa-Okamoto $\sigma$-form of Painlev\'e V:} {\small
\begin{align}
 & (t H_N'')^2=\left[t H_N'- H_N + H_N'(2N+\al+\la) + N \la \right]^2 \nonumber \\ & \hspace*{2cm} - 4(t
H_N'- H_N + \dN )\left[( H_N')^2+\la H_N'\right] \label{eq:JimboPV}
\end{align}}
\emph{with} $\dN := N(N+\al+\la)$.

In addition, $H_N(t)$ also admits a second-order non-linear difference representation, in terms of $N$ with $t$ fixed, which we call the discrete
$\sigma$-form. This is given by
\begin{align}  \label{eq:DiscreteSig}
&\left[\frac{N(N+\al)t+(\Delta^2H_N+t)(H_N-\delta_N)}{\Delta^2H_N+2N+\al+\la+t}\right]^2
+\la\:t\:\frac{N(N+\al)t+(\Delta^2H_N+t)(H_N-\delta_N)}{\Delta^2H_N+2N+\al+\la+t}\nn\\
& \hspace*{0.5cm}= \left[\delta_N-H_N
+\frac{N(N+\al)t+(\Delta^2H_N+t)(H_N-\delta_N)}{\Delta^2H_N+2N+\al+\la+t}\right]
(H_{N+1}-H_N)(H_N-H_{N-1}) \quad 
\end{align}
where $\Delta^2H_N:=H_{N-1}-H_{N+1}$. The initial conditions are
$H_1(t) = \frac{d}{dt}\log D_1(t,\la),\: H_2(t) = \frac{d}{dt}\log
D_2(t,\la)$ with $ D_1(t,\la) = \mu_0(t),\: D_2(t,\la) =
\mu_0(t)\mu_2(t)-\mu_1^2(t),$ and the moments are defined in
$(\ref{eq:MomentDefn}).$

{\bf Remark:}  \emph{We point out that Theorem 1 and the discrete $\sigma$-form in Theorem 2 are new, whilst the continuous $\sigma$-form in Theorem 2 was presented previously in \cite{ChenMckay}, and also in \cite{Ozipov} via different means (i.e., using an integrable-systems approach).}

{\bf Remark:}  \emph{We would also like to point
out that Painlev\'e equations first appeared in the early 1900's
through the work of Painlev\'e and his collaborators \cite{book}. In
the mid 1970's, Painlev\'e equations first appeared in
characterizing the correlation function of an Ising model through
the pioneering work of Barouch, McCoy, Tracy and Wu, see
\cite{Tracy}. The 1-particle reduced density matrix was shown in
1980 to satisfy a particular Painlev\'e V, see \cite{Jimbo}.  For a
recent review on this and other related problems in matrix
ensembles, see \cite{TW2011}. Another Painlev\'e V appeared in the
Hankel determinant associated with the ``time evolved" Jacobi
polynomials, see \cite{Basor}.}

\subsection{Proof of Theorems 1 and 2}

\subsubsection{Compatibility Conditions, Recurrence Coefficients and
Discrete Equations}



To get started, for the purpose of applying the ladder operator framework introduced
in Lemmas 1--3, first note that {\small \bq
\v(z,t)&:=&-\log w_{\rm dLag}(z,t)=-\al \log z-\la \log(z+t)+z , \nn\\
\v'(z,t)&=&-\frac{\al}{z}-\frac{\la}{z+t}+1\nn \eq 
with the derivative taken with respect to $z$. Therefore \bq
\frac{\v'(z,t)-\v'(y,t)}{z-y}&=&\frac{\al}{zy}+\frac{\la}{(z+t)(y+t)}.\nn
\eq}
Substituting the above into (\ref{eq:AnNewDefn}) and
(\ref{eq:BnNewDefn}), followed by integration by parts, we obtain
{\small \begin{align}
\label{eq:AnDefn}
A_n(z)&=\frac{1-R_n(t)}{z}+\frac{R_n(t)}{z+t}\\
B_n(z)&=-\frac{n+r_n(t)}{z}+\frac{r_n(t)}{z+t}
\end{align}}
where we have introduced the auxiliary variables:
{\small \begin{align}
R_n(t)&:=\frac{\la}{h_n}\int_{0}^{\infty}\frac{[P_n(x)]^2}{x+t}w_{{\rm dLag}}(x,t)dx\\
r_n(t)&:=\frac{\la}{h_{n-1}}\int_{0}^{\infty}\frac{P_n(x)P_{n-1}(x)}{x+t}w_{{\rm dLag}}(x,t)dx.
\label{eq:rnDefn}
\end{align}}
These auxilliary variables are particularly important in the subsequent derivations. The first major stage of the proof methodology involves relating these auxilliary variables to certain key quantities; primarily, 
the recurrence coefficients $\al_n$ and $\bt_n$, the coefficient $\textsf{p}_1(n)$ of $z^{n-1}$ in $P_n(z)$, as well as $\sum_{j=0}^{n-1} \:R_j$.  (Note that $\textsf{p}_1(n)$ also depends
on $t$ but we do not display this if there is no confusion.)  These relationships are established in the following:

\noindent {\bf Lemma 4} \emph{The recurrence coefficients $\al_n$ and $\bt_n$
relate to the auxiliary quantities $r_n$ and $R_n$ via:}
{\small \bq \al_n&=&2n+1+\al+\la-tR_n \label{eq:AlphanRelation} \\
\bt_n&=&\frac{1}{1-R_n}\left[r_n(2n+\al+\la)+\frac{r_n^2-\la
r_n}{R_n}+n(n+\al)\right]. \label{eq:BetanRelation}
\eq}
\emph{Furthermore,}
{\small \bq
t\sum_{j=0}^{n-1}R_j&=&n(n+\al+\la)-\bt_n-tr_n, \label{eq:SumRjRelation} \\
\textsf{p}_1(n)&=&-\bt_n-tr_n. \label{eq:p1Relation}
\eq}
{\bf Proof:}    We will derive (\ref{eq:AlphanRelation}), (\ref{eq:SumRjRelation}), (\ref{eq:p1Relation}), and (\ref{eq:BetanRelation}), in turn. These relationships are established based on the supplementary conditions, quoted in Lemma 2.  In particular, we start from  $(S_1)$. Equating the coefficients of $z^{-1}$ and $(z+t)^{-1}$, we obtain
the following difference equations relating $\al_n$ to $r_n$ and $R_n$:
%
%
{\small \begin{align}
\label{eq:S1Diff1}
-(2n+1+r_{n+1}+r_n)&=\al-\al_n(1-R_n)\\
r_{n+1}+r_n &= \la-R_n(t+\al_n).
\label{eq:S1Diff2}
\end{align}}
Summing these equations yields (\ref{eq:AlphanRelation}), the desired relation for $\alpha_n$.

To proceed further, we take note of
(\ref{eq:AnDefn})--(\ref{eq:rnDefn}), and derive
identities based on the supplementary condition $(S_2')$.
A straightforward but rather lengthy computation shows that the right-hand side of $(S_2')$ becomes
{\small \begin{align}
& B_n^2(z)+\v'(z)B_n(z)+\sum_{j=0}^{n-1}A_j(z) \nonumber \\
& \hspace*{2cm} = z^{-2}[(n+r_n)^2+\al(n+r_n)]\nn\\
& \hspace*{2cm} +z^{-1}\Big\{n-\sum_{j=0}^{n-1}R_j+r_n[\la-\al-t-2(n+r_n)]/t+(n-\la)/t\Big\}\nn\\
& \hspace*{2cm} +(z+t)^{-1}\Big\{\sum_{j=0}^{n-1}R_j+r_n[t+\al-\la+2(n+r_n)]/t-n\la/t\Big\}\nn\\
& \hspace*{2cm} +(z+t)^{-2}[r_n^2-\la r_n] . \nn
\end{align}}
Now focusing on $(S_2')$ as presented above, and equating the coefficients
of $z^{-2},$ $z^{-1},$ $(z+t)^{-1},$ $(z+t)^{-2}$,
give rise to the following difference equations involving $\beta_n$, $r_n$, $R_n$ and $\sum_{j=0}^{n-1} \:R_j$ :
%
%
{\small \begin{align}
\label{eq:S2Diff1} (n+r_n)^2+\al(n+r_n) =\bt_n(1-R_n)(1-R_{n-1})
\end{align}}
{\small \begin{align}
\small n-\sum_{j=0}^{n-1}R_j+\frac{r_n}{t}[\la-\al-t-2(n+r_n)]+\frac{n(\la-t)}{t} =
\frac{\bt_n}{t}\left[(1-R_{n-1})R_n+(1-R_{n-1})R_n\right]
\label{eq:S2Diff2a}
\end{align}}
{\small \begin{align}
\sum_{j=0}^{n-1}R_{j}+\frac{r_n}{t}[t+\al-\la+2(n+r_n)]-\frac{n\la}{t}
 =
-\frac{\bt_n}{t}\left[(1-R_n)R_{n-1}+(1-R_{n-1})R_n\right]
\label{eq:S2Diff2b}
\end{align}}
{\small \begin{align}
r_n^2-\la r_n =\bt_nR_nR_{n-1}.\quad \label{eq:S2Diff3}
\end{align}}
Observe that (\ref{eq:S2Diff2a}) and
(\ref{eq:S2Diff2b}) are equivalent. 

{\bf Remark:}  \emph{We shall see later (in Section \ref{sec:Final}) that (\ref{eq:S2Diff2a}), when combined with
certain identities, performs the sum $\sum_{j=0}^{n-1}R_j$
automatically in closed-form. This sum will provide an important
link between the logarithmic derivative of the Hankel determinant
with respect to $t$, and the quantities $\bt_n$ and $r_n$. }


From
(\ref{eq:S2Diff1}) and (\ref{eq:S2Diff3}) we find after a minor re-arrangement
{\small \bq \label{eq:RnBnCond}
\bt_n(R_{n}+R_{n-1})=\bt_n-n(n+\al)-r_n(\al+\la+2n).
\eq}
Now substituting (\ref{eq:S2Diff3}) and (\ref{eq:RnBnCond}) into either
(\ref{eq:S2Diff2a}) or (\ref{eq:S2Diff2b}) to eliminate $R_n$ and
$R_{n-1}$, we obtain (\ref{eq:SumRjRelation}), the desired relation for $R_n$.
 
In view of (\ref{eq:AlphanRelation}), we can also obtain an alternative representation for $\sum_{j=0}^{n-1}R_j,$ namely,
{\small \bq
t\sum_{j=0}^{n-1}R_j&=&n(n+\al+\la)-\sum_{j=0}^{n-1}\al_j
=n(n +\al+\la)+\textsf{p}_1(n) \label{eq:RSumb} .
\eq}
Comparing this with  (\ref{eq:SumRjRelation}) gives (\ref{eq:p1Relation}), the desired relation for $\textsf{p}_1(n)$.


Finally, by eliminating $R_{n-1}$ from (\ref{eq:RnBnCond}) and (\ref{eq:S2Diff3}), we obtain (\ref{eq:BetanRelation}), the desired relation for $\beta_n$.

\qed


\subsubsection{$t$ Evolution and a Connection to Painlev\'e V}
In the next stage of the development, the objective is to establish relationships between the auxilliary variables, $r_n$ and $R_n$, and \emph{derivatives} with respect to $t$ of the key quantities $\alpha_n$, $\beta_n$, $\textsf{p}_1(n)$, as well as $\log h_n$.  This, in turn, will allow us to establish a set of coupled Riccardi equations, which involve $r_n$ and $R_n$ and their derivatives.  Moreover, we will demonstrate that $R_n$, up to a simple linear fractional transformation, is the solution to a certain Painlev\'e V equation.


First, a straightforward computation shows that
{\small \begin{align} \label{eq:hnRel}
\frac{d}{dt}\log
h_n=R_n. 
\end{align}}
But, from (\ref{eq:betah}) and also (\ref{eq:S2Diff3}), it follows that
 {\small \bq
\frac{d\bt_n}{dt}&=&\bt_n(R_n-R_{n-1})\\
&=&\bt_n R_n-\frac{r_n^2-\la r_n}{R_n} \label{eq:BetaDiff} \; . \eq} 

Differentiating
{\small $$
0=\int_{0}^{\infty}x^{\al}(x+t)^{\la}\rme^{-x}P_{n}(x)P_{n-1}(x)dx
$$}
with respect to $t$ produces
{\small \begin{align}
0&=\la\int_{0}^{\infty}\:x^{\alpha}\:(x+t)^{\la-1}\rme^{-x}P_n(x)P_{n-1}(x)dx+
\int_{0}^{\infty}x^{\al}(x+t)^{\la}\rme^{-x}\left[\frac{d}{dt}\textsf{p}_1(n)\:x^{n-1}+...\right]P_{n-1}(x)dx\nn\\
&=\la\int_{0}^{\infty}\frac{P_{n-1}(x)P_{n}(x)}{x+t}w_{{\rm dLag}}(x,t)dx+h_{n-1}\:\frac{d}{dt}\textsf{p}_1(n),\nn
\end{align}}
resulting in
{\small \bq \label{eq:p1Diff}
\frac{d}{dt}\textsf{p}_1(n)&=&-r_n .
\eq}
Upon noting
(\ref{eq:p1Defn}), this implies
\bq \label{eq:AlphaDiff}
\frac{d\al_n}{dt}&=&r_{n+1}-r_{n} .
\eq
Now differentiating
(\ref{eq:p1Relation}) with respect to $t$, we find
{\small \bq
\frac{d}{dt}\textsf{p}_1(n)&=& -\frac{d\bt_n}{dt}-r_n-t\frac{dr_n}{dt} \nn  \; .
\eq}
The above result combined with (\ref{eq:p1Diff}) and  (\ref{eq:BetaDiff}) 
gives
{\small \bq
\frac{d\bt_n}{dt}=-t\frac{dr_n}{dt}=\bt_nR_n-\frac{r_n^2-\la
r_n}{R_n}.
\eq}
We now come to a key Lemma which gives the first order derivative of $r_n(t)$ and $R_n(t)$ with respect to $t,$ and where $n$
appears as a parameter.
\\
\\
{\bf Lemma 5} \emph{The auxiliary variables} $r_n$ \emph{and} $R_n$
\emph{satisfy the following coupled Riccatti equations,}
{\small \begin{align}
\label{eq:RicEqr} t\frac{dr_n}{dt}=\frac{r_n^2-\la
r_n}{R_n}-\frac{R_n}{1-R_n}\left[r_n(2n+\al+\la)+\frac{r_n^2-\la
r_n}{R_n}+n(n+\al)\right],
\end{align}}
\emph{and}
{\small \begin{align}
\label{eq:RicEqR} 2r_n=t\frac{dR_n}{dt}+\lambda-R_n\:(t+2n+\alpha+\lambda-t\:R_n).
\end{align}}
\emph{Furthermore,}
{\small $$
y(t)=y(t, n):=1-\frac{1}{1-R_n(t)},
$$}
\emph{satisfies the following second-order non-linear ordinary differential equation,}
\begin{align}
y''&=\frac{3y-1}{2y(y-1)}\:(y')^2- \frac{y'}{t}+\frac{(y-1)^2}{t^2}
\: \left(\frac{\al^2}{2}y-\frac{\la^2}{2y}\right)+\frac{(2n+1+\al+\la)\:y}{t}-\frac{y(y+1)}{2(y-1)},
\label{eq:PVyt}
\end{align}
\emph{which is recognized to be a Painlev\'e V,} {\small
$$P_V\left(\frac{\al^2}{2},-\frac{\lambda^2}{2},2n+1+\alpha+\lambda,-1/2\right).$$}

{\bf Proof: }Because (\ref{eq:BetanRelation}) expresses $\bt_n$ as a
quadratic in $r_n,$ we see that $r_n$ satisfies the Riccatti
equation (\ref{eq:RicEqr}). Eliminating $r_{n+1}$ from
(\ref{eq:S1Diff2}) and (\ref{eq:AlphaDiff}), and upon referring to
(\ref{eq:AlphanRelation}), we obtain (\ref{eq:RicEqR}). Next, we
simply substitute  $r_n(t)$ from (\ref{eq:RicEqR}) into
(\ref{eq:RicEqr}), to see that $R_n(t)$ satisfies a second-order
non-linear ordinary differential equation in $t$, in which $n$, $\al$, and $\la$ appear as
parameters. A further linear fractional change of variable {\small
$$ R_n(t)=1-\frac{1}{1-y(t)}\quad{\rm or}\quad
y(t) =1-\frac{1}{1-R_n(t)},
$$}
establishes that $y(t)$ satisfies the Painlev\'e V displayed in the Lemma.

\qed

{\bf Remark:} \emph{ This Painlev\'e V relationship for the auxilliary variable $R_n$ presents a new result which, along with faciliting the subsequent derivations, may also be of independent interest.}

\subsubsection{Connecting to the Hankel Determinant} \label{sec:Final}

Having developed the above relations for $r_n$, $R_n$, $\alpha_n$, $\beta_n$, and $h_n$, we are now in a position to employ those results to establish the two integral representations for the Hankel determinant of interest,
i.e., {\small
$$ D_N(t,\lambda)=
\det\left(\int_{0}^{\infty}x^{j+k-2}(x+t)^{\lambda}x^{\alpha}{\rm
e}^{-x}dx\right)_{1\leq j,k\leq N} \; ,$$} given in Theorems 1 and
2. To this end, noting (\ref{eq:hnRel}), an easy computation shows that {\small \bq
H_N(t)&:=&t\frac{d}{dt}\log D_N(t,\lambda)=t\frac{d}{dt}\sum_{j=0}^{N-1}\log h_j=t\sum_{j=0}^{N-1}R_j\nn\\
&=& N(N+\al+\la)-\bt_N-tr_N \label{eq:Hna} \\
&=& N(N+\al+\la)+\textsf{p}_1(N) \label{eq:Hnb}, \eq} where the last
two equations follow from (\ref{eq:SumRjRelation}) and
(\ref{eq:p1Relation}) of Lemma 4. Integrating (\ref{eq:Hna}) with
respect to $t$, while noting (\ref{eq:BetanRelation}),
(\ref{eq:RicEqR}) and $R_N(t)=1-1/(1-y(t)),$ we obtain the result
stated
in Theorem 1.

To obtain the second integral representation for $D_N(t,\lambda)$
stated in Theorem 2 (i.e., in terms of $H_N(t)$), we note that from
(\ref{eq:p1Diff}), (\ref{eq:Hna}), and (\ref{eq:Hnb}), we obtain
expressions for $\bt_N$ and $r_N$ in terms of $H_N$ and $H_N'$,
 \bq
\bt_N&=&N(N+\al+\la)+tH_N'-H_N \label{eq:BetanHn} \\
r_N &=& -H_N'. \eq What we need to do is to eliminate $R_N$ to find
a functional equation satisfied by $H_N,$ $H_N'$ and $H_N''.$ For
this purpose, we examine two quadratic equations satisfied by $R_N$,
one of which is simply a rearrangement of (\ref{eq:BetanRelation})
and reads \bq \label{eq:Frac1} \frac{r_N^2-\la r_N}{R_N}+\bt_N
R_N=\bt_N-r_N(2N+\al+\la)-N(N+\al). \eq The other follows from a
derivative of the first equation of (\ref{eq:BetanHn}) with respect
to $t$ and (\ref{eq:BetaDiff}), \bq \label{eq:Frac2} \bt_N
R_N-\frac{r_N^2-\la r_N}{R_N} &=& tH_N''. \eq Solving for $R_N$ and
$1/R_N$ from the linear system (\ref{eq:Frac1}) and
(\ref{eq:Frac2}), we find
\bq
2R_N&=&1+\frac{tH_N''-(2N+\la+\la)r_N-N(N+\al)}{tH_N'-H_N+N(N+\al+\la)} \label{eq:Rna} \\
\frac{2}{R_N}&=&\frac{-tH_N''+(t+2N+\al+\la)H_N'-H_N+N\la}{(H_N')^2+\la\:H_N'}
\label{eq:Rnb}, \eq where we have replaced $\beta_N$ and $r_N$ in
terms of $H_N,\;H_N' {\rm and\;} H_N''$ with (\ref{eq:BetanHn}). The
product (\ref{eq:Rna}) and (\ref{eq:Rnb}) gives us the desired
$\sigma$-form (\ref{eq:JimboPV}).
%

{\bf Remark:} \emph{
It is worth noting that with
$D_N(t,\la)=:t^{\dN}\tD_N,$
we find, after a little computation that $\tD_N$ satisfies the Toda
molecule equation \cite{Toda} \bq
\frac{d^2}{dt^2}\log\tD_N=\frac{\tD_{N+1}\tD_{N-1}}{\tD_N^2} . \eq }


Finally, we will compute the discrete $\sigma$-form in (\ref{eq:DiscreteSig}).  
For this, the proof is elementary. We start from  $H_N=t\sum_{j=0}^{N-1}R_j.$ It follows that
$H_{N+1}-H_{N}=t\:R_N,$ $H_N-H_{N-1}=t\:R_{N-1},$ and
\begin{align}
-\Delta^2H_{N}=t(R_N+R_{N-1}),
\label{eq:HN}
\end{align}
where $\Delta^2H_N:=H_{N-1}-H_{N+1}.$

The idea is to express $r_N,$ and $\beta_N$ in terms of $H_N,$ $H_{N\pm 1}$, $N$ and the parameters $\alpha$ and $t.$
Multiplying (\ref{eq:RnBnCond}) by $t,$ a little re-arrangement yields a linear equation in $\bt_N$ and $tr_N,$
\begin{align}
(t+\Delta^2H_N)\:\beta_N - (\alpha+\lambda+2N)\:t\:r_N=N(N+\alpha)t.
\label{eq:line1}
\end{align}
A little re-arrangement of (\ref{eq:SumRjRelation}) yields a further linear equation
\begin{align}
\beta_N+t\:r_N=N(N+\alpha+\lambda)-H_N.
\label{eq:line2}
\end{align}
Hence,
\begin{equation}
t\:r_N=\frac{(t+\Delta^2H_N)[N(N+\alpha+\lambda)-H_N]-N(N+\alpha)t}{2N+\alpha+\lambda+t+\Delta^2H_{N}}
\end{equation}
\begin{equation}
\beta_N=N(N+\alpha+\lambda)-H_N+
\frac{N(N+\alpha)t+[H_N-N(N+\alpha+\lambda)](t+\Delta^2H_N)}{2N+\alpha+\lambda+t+\Delta^2H_N}.
\end{equation}
Substituting these into (\ref{eq:S2Diff3}) yields the desired discrete $\sigma$-form.


\section{Concluding Remarks}

The objective of this article, whilst largely an expository review, was to demonstrate how the ladder operator approach can be applied to yield different characterizations of a certain Hankel determinant arising in
the information-theoretic study of MIMO communication systems (more specifically, when dealing with the moment generating function of the channel capacity).  The Hankel determinant of interest in this problem is generated from a certain deformed Laguerre weight, and for this determinant we evaluated two exact integral representations.  The first of these was described in terms of the solution to a certain non-linear differential equation, which appears new.  The second integral representation was described in two forms: the first form involving the solution to the $\sigma$-form of a particular Painlev\'{e} V differential
equation, which was reported previously in \cite{ChenMckay} and also \cite{Ozipov}, whilst the second form was stated in terms of a certain second-order non-linear difference equation, which also constitutes a new result.

\end{document}